\documentclass{PoS}
\usepackage{cite}
\newcommand{\be}{\begin{equation}}
\newcommand{\ee}{\end{equation}}
\title{Axial Anomaly and Transition Formfactors}

\ShortTitle{Axial Anomaly and Transition Formfactors}

\author{\speaker{Oleg Teryaev}
\\
        Bogoliubov Laboratory of Theoretical Physics, JINR, Dubna 141980, Russia\\
        E-mail: \email{teryaev@theor.jinr.ru}}

\author{Yaroslav Klopot \thanks{On leave from Bogolyubov Institute for Theoretical Physics, Kiev, Ukraine}
\\
        Bogoliubov Laboratory of Theoretical Physics, JINR, Dubna 141980, Russia\\
        E-mail: \email{klopot@theor.jinr.ru}}

\author{Armen Oganesian\\
Institute of Theoretical and Experimental Physics, Moscow 117218, Russia \\
        E-mail: \email{armen@itep.ru}}

\abstract{We study photon-meson transition formfactors of light mesons in the kinematics, where one photon is real and other is virtual. Dispersive approach to axial anomaly leads to  the anomaly sum rule. The absence of corrections to it allows us to get the relation between possible corrections to continuum and to lower states within QCD method which does not rely on factorization hypothesis. We show, relying on the recent data of the BaBar Collaboration, that the relative correction to continuum is quite small, and small
correction to continuum can dramatically change the pion formfactor. The same effect for $\eta$ meson is shown to be less pronounced.}

\FullConference{35th International Conference of High Energy Physics - ICHEP2010,\\
        July 22-28, 2010\\
        Paris France}

\begin{document}

\section{Introduction}
Theoretical study of the photon-meson transition formfactor has a long history and have yielded several important results. It is well known, that the transition formfactor  of the pion $F_{\pi\gamma}(k^2,q^2)$ into two \textit{real} photons ($k^2=0,q^2=0$) is governed by axial anomaly \cite{Bell:1969ts}:
$F_{\pi\gamma}(0,0)=1/(2\sqrt{2}\pi^2 f_\pi), \;$ $f_\pi=130.7 \;MeV$.
On the other hand,  in the kinematical region where one photon is real ($k^2=0$) and other is virtual ($q^2=-Q^2<0$), factorization approach to perturbative quantum chromodynamics (pQCD) for exclusive process in the leading order in the strong coupling constant predicts
\cite{Lepage:1980fj,Brodsky:1981rp}:
$F_{\pi\gamma}(Q^{2})=\frac{\sqrt{2}f_\pi}{3Q^{2}}
\int_{0}^{1}dx\frac{\varphi_{\pi}\left( x,Q^2
\right)}{x}+\mathcal{O}(1/Q^4) \;,
$
where $\varphi_{\pi}(x)$ is a pion
distribution amplitude (DA). The pion DA depends on the
renormalization scale and at large $Q^2$ asymptotically acquires form $\varphi_{\pi}^{\mathrm{asymp}}\left(
x\right)=6x\left( 1-x \right)$. This leads to asymptotic
behaviour for the pion transition formfactor at $Q^2 \to \infty$:

\begin{equation}\label{AsLargeQ}
F_{\pi\gamma}^{\mathrm{asymp}}(Q^{2})=\frac{\sqrt{2} f_{\pi}}{Q^{2}}+\mathcal{O}(1/Q^4) \;.%
\end{equation}

Recent measurements of $ F_{\pi\gamma}(Q^{2})$ at  $4<Q^2<40$
$GeV^2$ by BaBar Collaboration \cite{Aubert:2009mc} showed very unexpected results: although at $Q^2<10$ $GeV^2$ the data show good agreement with previous experiments and pQCD predicted behaviour, at larger virtualities transition formfactor continues to grow strongly exceeding the predicted asymptotics (\ref{AsLargeQ}). This disagreement leads to to the question of pQCD factorization validity. Recently, there were proposed several approaches to explain such unusual behaviour of $F_{\pi\gamma}(Q^2)$ \cite{Dorokhov:2009dg,Radyushkin:2009zg,Polyakov:2009je,Chernyak:2009dj}, in particular,
questioning pQCD factorization.

The aim of our work is to study the meson-photon transition formfactors for the case of virtual photon using the anomaly sum rule. This generalizes the usual application of anomaly which provides the boundary condition in the limit of two real photons only. Our (non-perturbative) QCD method does not imply the QCD
factorization and is valid even if the QCD factorization is broken.
Using axial anomaly in the dispersive approach we get the exact relations between possible corrections to
lower states and continuum providing a possibility of relatively
large corrections to the lower states.

\section{Anomaly sum rule and quark-hadron duality}
The anomaly sum rule can be derived using the dispersive approach to axial anomaly. Let us introduce notations and briefly remind the derivation of the anomaly sum rule (see \cite{Horejsi:1994aj}).
The VVA triangle graph amplitude
\be T_{\alpha \mu\nu}(k,q)=\int
d^4 x d^4 y e^{(ikx+iqy)} \langle 0|T\{ J^5_\alpha(0) J_\mu (x)
J_\nu(y) \}|0\rangle \label{VVA} \ee
contains axial current $J^5_\alpha=(\bar{u}\gamma_5 \gamma_\alpha u -\bar{d}\gamma_5\gamma_\alpha d)$ and two
vector currents $J_\mu = ((2/3)\bar{u}\gamma_\mu u -(1/3)\bar{d}\gamma_\mu d)$;  $k,q$ are momenta of photons.
This amplitude can be presented as a tensor decomposition

\begin{eqnarray}
\label{eq1} \nonumber T_{\alpha \mu \nu} (k,q) & = & F_{1} \;
\varepsilon_{\alpha \mu \nu \rho} k^{\rho} + F_{2} \;
\varepsilon_{\alpha \mu \nu \rho} q^{\rho}
 + \; \; F_{3} \; q_{\nu} \varepsilon_{\alpha \mu \rho \sigma}
k^{\rho} q^{\sigma} + F_{4} \; q_{\nu} \varepsilon_{\alpha \mu
\rho \sigma} k^{\rho}
q^{\sigma}\\
 & & \;\;\;\;+ \; \; F_{5} \; k_{\mu} \varepsilon_{\alpha \nu
\rho \sigma} k^{\rho} q^{\sigma} + F_{6} \; q_{\mu}
\varepsilon_{\alpha \nu \rho \sigma} k^{\rho} q^{\sigma} \;,
\end{eqnarray}
where the coefficients $F_{j} = F_{j}(k^{2}, q^{2}, p^{2}; m^{2})$, $p
= k+q$, $j = 1, \dots ,6$ are the corresponding Lorentz invariant
amplitudes.
Writing the unsubtracted dispersion relations for the formfactors
one gets the finite subtraction for axial current divergence resulting
in the anomaly sum rule which for the kinematical configuration we are interested in ($k^{2} =
0$, $q^{2} \not= 0$) takes the form \cite{Horejsi:1994aj}:
\begin{equation}
\label{ASR} \int_{4m^{2}}^{\infty} A_{3a}(t;q^{2},m^{2}) dt =
\frac{1}{2\pi}, \;\;\;\; A_{3a}\equiv\frac{1}{2}Im (F_3-F_6).
\end{equation}
This anomaly sum rule (ASR) relation holds for an
arbitrary quark mass $m$ and for any $q^{2}$ in the considered
region. Another important property of the above relation is
absence of any $\alpha_s$ corrections to the integral. Moreover,
it is expected that it does not have any nonperturbative
corrections too ('t Hooft's principle).

The pion transition formfactor  $F_{\pi\gamma}$, which is defined from the matrix element
\be \int d^{4}x e^{ikx} \langle \pi^0(p)|T\{J_\mu (x) J_\nu(0)
\}|0\rangle = \epsilon_{\mu\nu\rho\sigma}k^\rho q^\sigma
F_{\pi\gamma} \; \ee

enters the tree-point correlation function $T_{\alpha
\mu\nu}(k,q)$:
\begin{equation} T_{\alpha
\mu\nu}(k,q)=\frac{i\sqrt{2}f_\pi}{p^2-m_{\pi}^2}p_\alpha k^\rho
q^\sigma \epsilon_{\mu\nu\rho\sigma} F_{\pi\gamma}+ (higher\;
states), \;\;\;\;\; \langle 0|J^5_\alpha(0) |\pi^0(p)\rangle=
i\sqrt{2}p_\alpha f_\pi.
\end{equation}

Therefore, the pion and higher states contribute the function $A_{3a}$ as follows:
\be A_{3a}= \sqrt{2} f_\pi \pi
F_{\pi\gamma}(Q^2) \delta (s-m_\pi^2)+ (higher\; states),
\ee

so from (\ref{AsLargeQ}) the pion and other contributions to ASR (\ref{ASR}) are

\be \frac{1}{2\pi}= \frac{2\pi f_\pi^2}{Q^2}+(higher\; states).
\ee

We see, that at $Q^2\neq0$ anomaly sum rule (\ref{ASR}) cannot be
saturated by pion contribution only due to $1/Q^2$ behavior, so we
need to consider higher states. The other contributions are provided by axial mesons, the lightest of which
is the $a_1(1260)$ meson. In fact, the  contribution  of longitudinally
polarized  $a_1$ is given by the similar equation to
(\ref{AsLargeQ}) at large $Q^2$.
Actually, the same is  true for all the
higher axial mesons and mesons with higher spin. So, for the case $Q^2\neq 0$ the anomaly relation
(\ref{ASR}) cannot be explained in terms of any finite number of
mesons due to the fact that all transition formfactors are
decreasing functions. That is why we conclude that only
\textit{infinite} number of higher states can saturate anomaly
sum rule and therefore at $Q^2\neq0$ the \textit{axial anomaly is
a collective effect of meson spectrum} in contrast with
the case of two real photons $Q^2=0$, where the anomaly sum rule
is saturated by pion contribution only. Let us note that this
conclusion does not depend on  choice of meson distribution amplitudes.

According to the quark-hadron duality, in the model "$\pi^0$+continuum", the spectral density $A_{3a}$ can be written as:
\begin{equation} \label{spec}
A_{3a}\left(s,Q^2\right)= \sqrt{2}\pi f_{\pi}\delta(s-m_\pi^2)
F_{\pi \gamma}\left(Q^2\right)\ +
A^{QCD}_{3a}\theta(s-s_0),
\end{equation}

where $A^{QCD}_{3a}\theta(s-s_0)$ is a continuum contribution with a threshold $s_0=0.7$ $GeV^2$. One-loop perturbative theory  calculation gives a simple result for $A^{QCD}_{3a}$:
$A^{QCD}_{3a}(s,Q^2) = \frac{1}{2\pi}{{Q^2}\over{(s+Q^2)^2}}.$
Using ASR relation (\ref{ASR}) one can easily obtain the expression for $F_{\pi\gamma}(Q^2)$, which coincides with the interpolation formula proposed by S.~Brodsky and G.~Lepage \cite{Brodsky:1981rp} $(s_0=4\pi^2 f^2_\pi)$
\be \label{qq}
F_{\pi\gamma}(Q^2)=\frac{1}{2\sqrt{2}\pi^2
f_\pi}\frac{s_0}{s_0+Q^2}. \ee
If we take into account the next contributing state ($a_1$ meson), the ASR can be written as follows:
\be \label{pi+a1+cont} \frac{1}{2\pi}=\sqrt{2}\pi f_\pi
F_{\pi\gamma}(Q^2)+
I_{a_1}+\frac{1}{2\pi}\int_{s_1}^{\infty}ds\frac{Q^2}{(s+Q^2)^2}
\;, \ee
so the $a_1$ contribution  $I_{a_1}$ can be estimated as (Eq. (\ref{qq}) is used):
\be I_{a_1}=\frac{1}{2\pi}Q^2\frac{s_1-s_0}{(s_1+Q^2)(s_0+Q^2)}
\;. \ee

The plot for contributions of pion, $a_1$ meson and continuum is shown in Fig.1.
The figure illustrates the anomaly collective effect: indeed, the
contribution of infinite number of higher resonances (continuum
contribution) dominates starting from relatively small $Q^2\simeq
1.5$ $GeV^2$.

\begin{figure} \center
  \includegraphics[width=0.7\textwidth, height=0.4\textwidth]{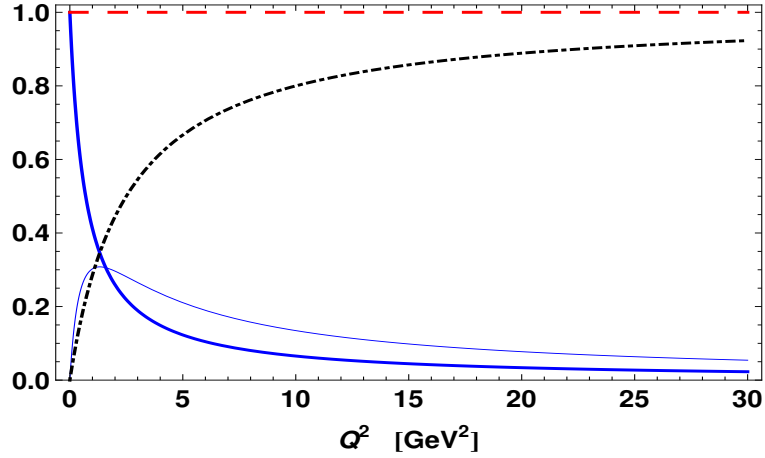}\\
  \caption{\textit{(Colour online).} Relative contributions of $\pi^0$ (thick blue curve), $a_1$ (thin blue
  curve) mesons (intervals of duality are $0.7$ $GeV^2$ and $1.8$ $GeV^2$ respectively), and continuum (
  dot-dashed black curve) (continuum threshold is $s_1=2.5$ $GeV^2$) to the anomaly sum rule (dashed red
  line).}\label{fig1}
\end{figure}

\section{Interplay between corrections and experimental data}
As we stressed above, anomaly sum rule  is an exact relation, i.e. $\int_{0}^{\infty} A_{3a}(s;Q^{2})ds$ does not acquire any corrections. Nevertheless, the continuum contribution $
I_{cont}=\int_{s_i}^{\infty} A_{3a}(s;Q^{2})ds $
may have perturbative as well as power corrections.
Note that  the two-loop corrections to the whole triangle graph
were found to be zero \cite{Jegerlehner:2005fs} implying the zero corrections
to all spectral densities. Therefore, the model of the corrections to continuum discussed
below should rather correspond to some non-perturbative corrections.
Let us first consider the contributions of local condensates. Naively, they should
strongly decrease with $Q^2$ compensating the mass dimension of gluon
(as quark one is suppressed even more)
condensate. However the 't Hooft's principle requires
(see \cite{Horejsi:1994aj}, Section 4)
the rapid decrease of the corrections with Borel parameter $M^2$ (related to $s$) so that
the power of $Q^2$ in the denominator may be not so large.
In reality the actual calculations do not satisfy
this property and the situation  may be
improved  by the use of non-local condensates
(see \cite{Horejsi:1994aj} and references therein). Another
possibility is other non-perturbative contributions,
like instanton-induced ones. So we assume the appearance of such
corrections in what follows modelling the corrections to continuum.
For the model ``$\pi^0$+continuum''
\be
\frac{1}{2\pi}=I_{\pi}+I_{cont}, \;\;\;\;\;I^0_{\pi}=\sqrt{2}\pi f_\pi
F^0_{\pi\gamma}(Q^2)= \frac{1}{2\pi}\frac{s_0}{s_0+Q^2},\;\;\;\;\;
 I^0_{cont}= \frac{1}{2\pi}\frac{Q^2}{s_0+Q^2}
\ee
the ASR requires the relation between corrections: $\delta I_\pi=-\delta I_{cont}$. However, the relative correction to pion is enhanced as compared to the relative correction to continuum by factor $Q^2/s_0$: \be \label{ratio}
R_\pi=\delta I_\pi/I^0_\pi= (\delta I_{cont}/I^0_{cont})\frac{Q^2}{s_0}.
\ee
which leads to the situation, where the leading power  correction to
continuum preserving its asymptotics results in a substantial (of
the order of the main term $I^0_\pi$) contribution to the pion
state changing the pion formfactor asymptotics at large $Q^2$. Supposing the correction to continuum to be $\delta I_{cont}=-c s_0
\frac{\ln(Q^2/s_0)+b}{Q^2}$, we can fit the parameters $b,c$ using data of BaBar collaboration for $F_{\pi\gamma}$: $b=-2.74,\;\; c=0.045$.

Estimating the relation between the corrections to continuum and to $\eta$ meson in the same way, we find that the enhancement is more than 3 times smaller than the one for $\pi^0$ (cf. V.~Druzhinin, these Proceedings):
\be R_{\eta}=\delta I_\eta/I^0_\eta \simeq (\delta I_{cont}/I^0_{cont})\frac{Q^2}{s_0^\eta} = s_0^\pi/s_0^\eta\simeq 0.36 R_{\pi}, \;\;\;\;\; s_0^\pi=0.7 GeV^2, \;\;s_0^\eta= 2.5 GeV^2. \ee

We thank V.~M.~Braun, V.~P.~Druzhinin, B.~L.~Ioffe, A. Khodjamirian, S.~V.~Mikhailov, S.~Narison  and
A.~V.~Radyushkin for useful discussions and comments. This  work
was supported in part by RFBR (grants 09-02-00732, 09-02-01149, 08-02-01003),
by the funds from EC to the project ``Study of the Strong Interacting
Matter''  under contract No. R113-CT-2004-506078 and
by  CRDF  Project  RUP2-2961-MO-09.

\end{document}